\documentclass[aps,pre,twocolumn,superscriptaddress]{revtex4}
\usepackage{amssymb}
\usepackage{amsmath}
\usepackage{times}
\usepackage{color}
\usepackage{graphicx}

\begin{document}

\title{Exploring agent interaction patterns in the comment sections of fake and real news}

\author{Kailun Zhu}
\affiliation{Institute of Cyberspace Security, Zhejiang University of Technology, Hangzhou, 310023, China}
\affiliation{Binjiang Institute of Artificial Intelligence, Zhejiang University of Technology, Hangzhou, 310056, China}

\author{Songtao Peng}
\affiliation{Institute of Cyberspace Security, Zhejiang University of Technology, Hangzhou, 310023, China}
\affiliation{Binjiang Institute of Artificial Intelligence, Zhejiang University of Technology, Hangzhou, 310056, China}

\author{Jiaqi Nie}
\affiliation{Institute of Cyberspace Security, Zhejiang University of Technology, Hangzhou, 310023, China}
\affiliation{Binjiang Institute of Artificial Intelligence, Zhejiang University of Technology, Hangzhou, 310056, China}

\author{Zhongyuan Ruan}
\email{zyuan.ruan@gmail.com}
\affiliation{Institute of Cyberspace Security, Zhejiang University of Technology, Hangzhou, 310023, China}
\affiliation{Binjiang Institute of Artificial Intelligence, Zhejiang University of Technology, Hangzhou, 310056, China}

\author{Shanqing Yu}
\email{yushanqing@zjut.edu.cn}
\affiliation{Institute of Cyberspace Security, Zhejiang University of Technology, Hangzhou, 310023, China}
\affiliation{Binjiang Institute of Artificial Intelligence, Zhejiang University of Technology, Hangzhou, 310056, China}

\author{Qi Xuan}
\email{xuanqi@zjut.edu.cn}
\affiliation{Institute of Cyberspace Security, Zhejiang University of Technology, Hangzhou, 310023, China}
\affiliation{Binjiang Institute of Artificial Intelligence, Zhejiang University of Technology, Hangzhou, 310056, China}

\begin{abstract}
User comments on social media have been recognized as a crucial factor in distinguishing between fake and real news, with many studies focusing on the textual content of user reactions. However, the interactions among agents in the comment sections for fake and real news have not been fully explored. In this study, we analyze a dataset comprising both fake and real news from Reddit to investigate agent interaction patterns, considering both the network structure and the sentiment of the nodes. Our main findings reveal that, compared to fake news, where users generate more negative sentiment, real news tend to elicit more neutral and positive sentiments. Additionally, nodes with similar sentiments cluster together more tightly than anticipated. From a dynamic perspective, we found that the sentiment distribution among nodes stabilizes early and remains stable over time. These findings have both theoretical and practical implications, particularly for the early detection of real and fake news within social networks.
\end{abstract}

\maketitle

\section{Introduction}

Online social platforms have emerged as predominant mediums for information circulation and user interaction. However, due to the low barriers to entry on these platforms, the spread of false information has become rampant, posing a significant challenge to social stability \cite{Lazer2018,Olan2024,Woolley2016}. For instance, a survey on COVID-19 vaccine acceptance found that misinformation about adverse side effects intensified hesitancy toward the vaccine \cite{Lazarus2023,Loomba2021}. Another example is the prevalence of political fake news, which often carries distinct biases and fuels social division \cite{Ho2022,Bovet2019,Pennycook2021}. During the 2016 U.S. elections, fabricated stories were widely distributed via social networks, impairing voters' rational judgment and endangering democracy. In economics, fake news characterized by hyperbole and falsehoods can mislead investors, triggering bouts of panic buying or selling and resulting in market instability \cite{Velichety2022}.

Therefore, distinguishing between fake and real information on online social networks has become an urgent task. In recent years, this issue has received considerable attention \cite{Vosoughi2018,Xuan2019,Juul2021,Kumar2018,Shu2019,Ruan2018}. Some early studies have concentrated on the textual or user characteristics of fake news \cite{Mukherjee2013,Kumar2016,Horne2017,Ruan2020,Ferrara2016,Yan2023}. For example, it was found that the titles of fake news are generally longer, use fewer stopwords, and contain more proper nouns and verb phrases, aiming to convey as much information in the title as possible. In contrast, the body content of real news articles tends to be short, repetitive, and less informative \cite{Horne2017}. Other studies have pointed out that many users involved in spreading fake news are ``throwaway" or automated accounts, such as social bots \cite{Ruan2020,Ferrara2016,Yan2023}, which is a distinguishing characteristic from real news. Additionally, several recent studies have focused on the propagation patterns of true and false information on social platforms \cite{Raponi2022,Djenouri2023,Vosoughi2018,Zannettou2017}, showing that false information propagates deeper and faster than true information, both on a single platform and across multiple platforms \cite{Vosoughi2018,Zannettou2017}.

Another common approach to identifying false information is to explore how users respond to or discuss fake news \cite{Setty2020,Ruchansky2017,Guo2019,Hamed2023}. People may express their opinions and interplay with each other in the comment sections of social platforms \cite{Medvedev2019}, which has been recognized as an important factor in detecting fake news. For example, some machine learning models have considered both the textual content and emotional characteristics of comments as key features \cite{Setty2020,Guo2019}. However, it remains unclear how users may interact with each other in the comment spaces of fake versus real news. Do users behave differently or similarly in these two scenarios? So far, a comprehensive study of this issue is still lacking. In this paper, we aim to fill this gap by investigating the interaction patterns of agents who participate in the discussions of online news. Specifically, we take into account the social interaction structures and sentiment polarities of the comments (or users), examining the differences/similarities in the comment trees for fake and real news, as well as in the user networks (constructed from the comment trees, see Sec. III).

This paper is organized as follows. In Sec. II, we introduce the data used in this paper. In Sec. III, we present the results based on the comment trees and user networks. Finally, we provide a discussion in Sec. IV.

\section{Dataset}
In this study, we focus on Reddit, a widely recognized social platform for news distribution and discussion. Reddit hosts millions of subforums covering a vast array of topics, including science, technology, entertainment, and everyday life. Moreover, Reddit users typically remain anonymous by using pseudonyms or screen names, which allows for greater expressive freedom. This anonymity enables users to openly post news, videos, images, or texts and engage in discussions across various threads. Due to its topical variety and user anonymity, Reddit has become extensively used for research purposes \cite{Medvedev2019,Cinelli2021,Curiskis2020}.

\begin{figure*}
\includegraphics[width=0.8\linewidth]{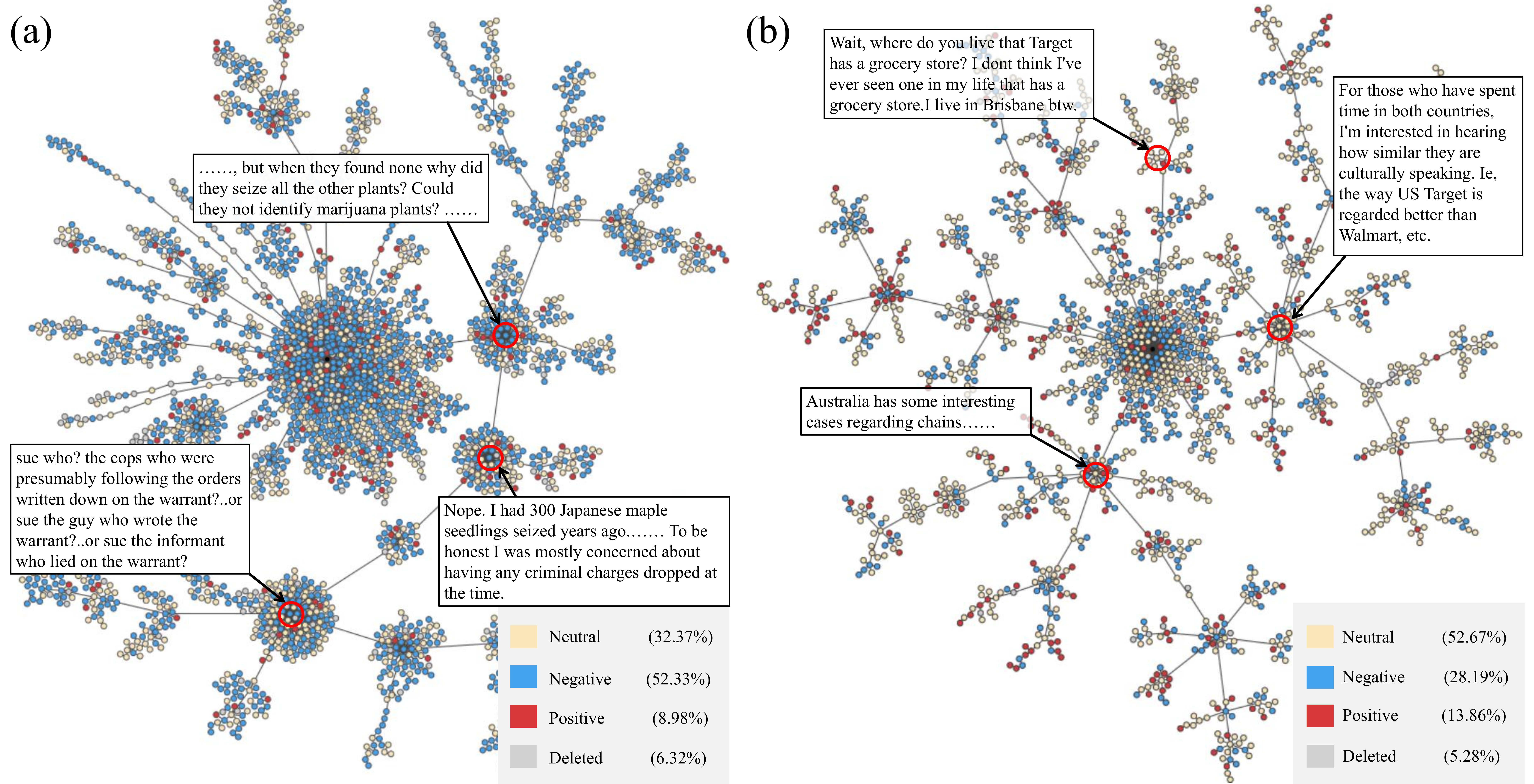} \caption{Examples of comment trees for (a) fake news, (b) real news. The yellow, blue and red colors represent neutral, negative, and positive sentiments of the comment contents, respectively. The gray color indicates that the contents have been deleted from the database. Therefore, the sentiments of these comments are unknown.} \label{Fig:1}
\end{figure*}

Our aim in this paper is to investigate the agent interaction patterns for fake and real news on Reddit. To achieve this, we utilize a dataset compiled by Setty et al., which includes statements labeled by \url{Snopes.com}, \url{Politifact.com}, and \url{Emergent.com} \cite{Setty2020}. This dataset is annotated with ``true'' and ``false'' labels to indicate the veracity of the content \cite{Website1}. Each post is associated with a unique link ID, allowing us to crawl the complete original text and comments using Reddit's official API. We further exclude posts with fewer than $20$ comments and preprocess the data by eliminating redundant URLs and symbolic emoticons, thereby reducing noise within the dataset. Moreover, we note that in this dataset, a few real posts are associated with an extremely high number of comments compared to fake news. To ensure comparability, we additionally exclude the 8 real posts with the highest number of comments, so that the maximum number of comments for both fake and real news posts is under $5,000$. Ultimately, we obtain $659$ posts labeled as ``real” with a total of $93,071$ comments, and $235$ posts labeled as ``fake” with a total of $43,547$ comments. 

Next, we employ a black-box tool named pysentimiento \cite{Perez2021}, a pre-trained BERT-based model that processes textual content and generates output \cite{Website2}, to determine the sentiment polarity of each comment. The model outputs probability values for three sentiment polarities: positive, neutral, and negative. Positive comments include those expressing known positive emotions (such as happiness, joy) and attitudes like recognition, hope, and belief; negative comments include those containing negative emotions (such as fear, doubt) and sentiments like abuse and disapproval; and neutral comments are those that describe events objectively without any obvious emotions. These values are ranked in descending order, and the highest probability indicates the corresponding sentiment polarity of the input text. By this way, we could assign each comment a discrete value $x$, where $x\in[-1,0,1]$ represents negative, neutral, and positive sentiment of the comment, respectively. Moreover, this model can also assess finer-grained emotions in the comments, encompassing six basic emotions: joy, surprise, sadness, disgust, anger, and fear, which can be considered as a refinement of sentiment polarity.

\section{Results}

\subsection{Static characteristics of the comment trees}
We treat the original posts and their associated comments as nodes, where each post acts as the source node. When node A comments on node B, a link is formed between them, resulting in the formation of numerous tree-like comment networks. These comment trees consist of nodes expressing different sentiments, with certain groups forming densely for both fake and real news (figure 1). By comparing real news with fake news, two interesting phenomena can be observed:  (i) For fake news, there is a possible trend that groups (defined as star subnetworks where the central node has a degree of at least $m=3$) are more prevalent within comment trees. The results are robust across different choices of $m$, provided it is not too large. (ii) Compared to fake news, which tends to generate more negative sentiment comments, real news is more likely to elicit neutral and positive responses. Moreover, comments with similar sentiment polarity tend to cluster together more tightly than anticipated.

To illustrate the first phenomenon, we count the number of nodes with degrees $k\ge m=3$ (referred to as backbone nodes) in each comment tree. The distribution is long-tailed, with only a few large trees containing many such nodes. We calculate the average number of the backbone nodes per tree (denoted as $\bar n_{\ge 3}$), finding that $\bar n_{\ge 3}=26.17$ for fake news and $\bar n_{\ge 3}=20.31$ for real news. This provides a first indication that fake news may generate more extensive discussions, consistent with the findings in \cite{Vosoughi2018}. However, a permutation test on these results yields a p-value of $0.19$, which is higher than the conventional threshold of statistical significance (typically $0.05$). While this suggests the results are not statistically significant, it is acceptable given the limited amount of data in our analysis, warranting further investigation with a larger dataset.

The second phenomenon can be observed by analyzing the distribution of sentiment polarities for nodes in the comment trees. For each tree $i$, we calculate the proportions of nodes with negative, neutral, and positive sentiment, denoted as $\eta_{-1}^i$,  $\eta_{0}^i$, and $\eta_{1}^i$ respectively. In figure 2a, we plot the distributions of  $\eta_{-1}^i$,  $\eta_{0}^i$, and $\eta_{1}^i$ for both fake and real news. Our findings indicate that the values of $\eta_{-1}^i$ for fake news trees are generally larger compared to those for real news, suggesting that fake news induces more negative sentiment. However, the distributions of $\eta_{0}^i$ and $\eta_{1}^i$ show a higher proportion of neutral and positive sentiments in real news, indicating that real news generates more balanced and positive discussions compared to fake news.

To reinforce this analysis, we take into account the finer-grained emotions expressed in the comments. We particularly focus on the emotions of joy, disgust, anger, fear, and sadness, where joy typically represents positive sentiment, while the others generally indicate negative sentiment (we here exclude surprise since it can be either positive or negative \cite{Plutchik2001}). Instead of the three-category classification, we calculate the proportion of nodes in each emotional state, $\eta_x$, for each tree, where $x \in \{joy, disgust, anger, fear, sadness\}$. Figure 2b shows the average and standard deviation of $\eta_x$ for both fake and real news. The results indicate that fake news tends to elicit higher average proportions of nodes expressing emotions such as disgust and anger compared to real news (the differences in fear and sadness however is not prominent), whereas real news exhibits a greater proportion of nodes expressing joy.

\begin{figure}
\includegraphics[width=1.0\linewidth]{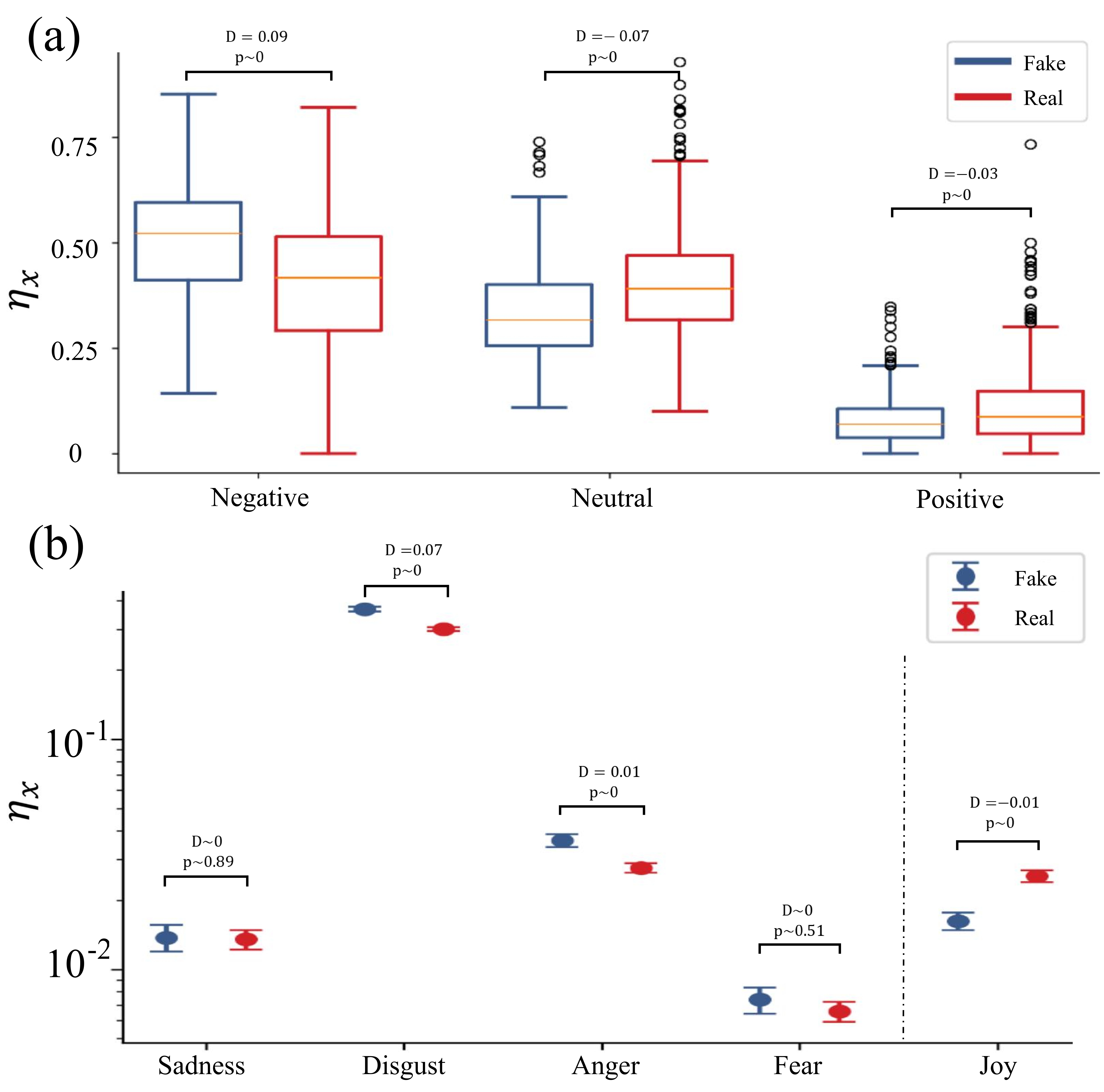} \caption{(a) Box plots of the distributions of $\eta_{x}^i$, where $x\in\{-1,0,1\}$. $\eta_{x}^i$ denotes the proportion of nodes with polarity $x$ in comment tree $i$. (b) Mean and variance of $\eta_{x}^i$, where $x$ represents finer-grained emotions, including joy, disgust, anger, fear, and sadness. A permutation test was conducted on these results. The p-value of less than $0.05$ indicates a significant difference in the means between the two sets.} \label{Fig:2}
\end{figure}

It should be noted that the dominant distribution of one type of sentiment may naturally lead to the clustering of nodes with that sentiment. However, the situation is more complex here. To illustrate this, we consider the correlation between nodes with different polarities. We introduce a quantity $r_{x,y}=l_{x,y}/l_x$ ($x,y\in\{-1,0,1\}$), which denotes the average fraction of neighbors with sentiment polarity $y$ pointing to a node with polarity $x$. Here, $l_{x,y}$ is the number of links pointing from nodes with polarity $y$ to nodes with polarity $x$, and $l_x=\sum_y l_{x,y}$ is the total number of neighbors pointing to nodes with sentiment polarity $x$. Figure 3a shows the average polarity distribution among the neighbors for a node with negative, neutral, and positive polarity in the network consisting of all fake news trees. As a comparison, we construct a null model in which the underlying network structure remains unchanged, but the sentiment polarities of nodes are randomly shuffled. The results are shown in figure 3b. By comparing the above two subfigures, we see that the fraction of nodes with either negative, neutral, or positive polarity around a node of the same polarity in the real case is higher than in the null model (the t-test results are displayed in figure 3e). For example, for negative sentiment, the proportion increases by a relative rate of $27\%$, from $0.51$ to $0.65$. These results demonstrate that nodes with similar polarity tend to cluster more tightly than expected. A similar phenomenon can also be observed in the scenario of real news (figure 3c,d). 

\begin{figure}
\includegraphics[width=1.0\linewidth]{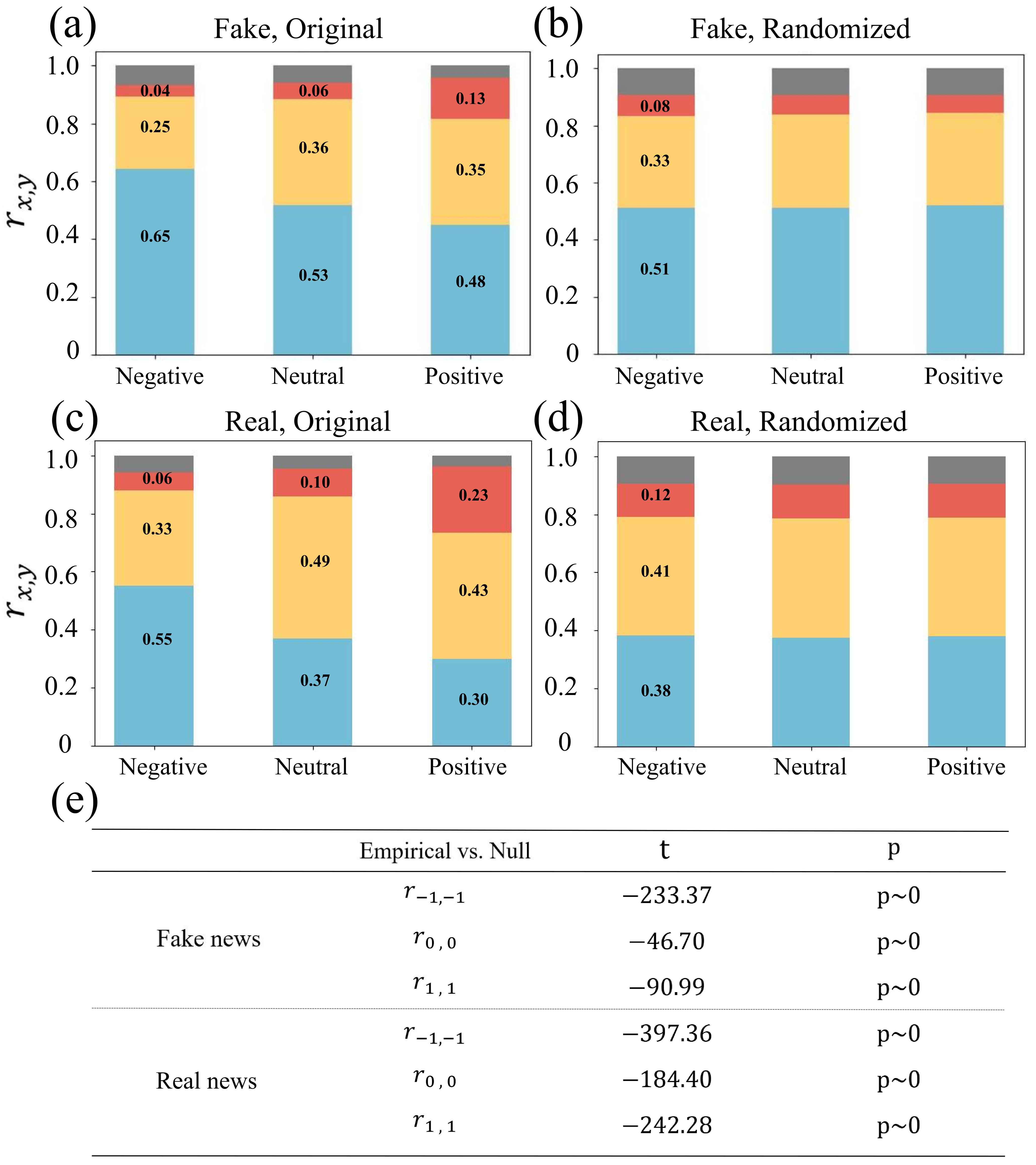} \caption{(a,b) Average fraction of neighbors with sentiment polarity $y\in\{-1,0,1\}$ pointing to a node with polarity $x\in\{-1,0,1\}$, $r_{x,y}$, for both fake and real news. (c,d) $r_{x,y}$ in the corresponding null models, where the sentiment polarities of nodes in comment trees are randomly shuffled (results averaged over $100$ realizations). The colors blue, yellow, and red represent negative, neutral, and positive sentiments, respectively, while gray corresponds to unknown sentiment. (e) One-sample t-test statistics comparing the empirical distribution to the null models.} \label{Fig:3}
\end{figure}

\subsection{Dynamic characteristics of the comment trees}

In this section, we focus on the evolutionary characteristics of comment trees using the information of the creation moment of each comment. Consider a post that contains a total of $N$ comments. Their commenting times can be recorded in an increasing order as $\{t_1, t_2, ..., t_N\}$, where $t_1$ and $t_N$ represent the times of the first and last comments, respectively, and the lifespan of the post is $t_{N} - t_{1}$. Note that the timestamps are precise to the second, making overlaps rare. In this way, we can readily evaluate the tree size at any time $t\le t_N$ by counting the number of comments made up to time $t$.

We then examine how the distribution of node sentiments evolves over time. As time progresses, the size of a tree increases. We divide this process into five stages: the first stage (S1) corresponds to the point when the tree reaches $20\%$ of its total size, the second stage (S2) covers the growth from $20\%$ to $40\%$, and so on. For convenience, we regard the original post as stage 0. Figure 4a,b shows the fraction of nodes with different sentiment polarities at each stage for all comment trees of fake news and real news. We find that the sentiment distribution in the later stages mirrors that of the initial stages as time advances, in other words, the sentiment distribution stabilizes in the very early stages of evolution. A finer-grained emotion analysis, as shown in figure 4c,d, reveals similar results, particularly for emotions with relatively larger proportions, such as disgust, anger, and joy. One possible reason for this is that early participants rapidly establish an emotionally charged comment environment, which influences later users to align with the early emotional patterns. To support this conjecture, we investigate the connection tendencies of nodes at each stage and find that nearly half of all nodes establish links with those from stage $1$ (for both fake and real news), suggesting the potentially far-reaching influence of early participants in shaping the overall sentiment dynamics.

\begin{figure}
\includegraphics[width=1.0\linewidth]{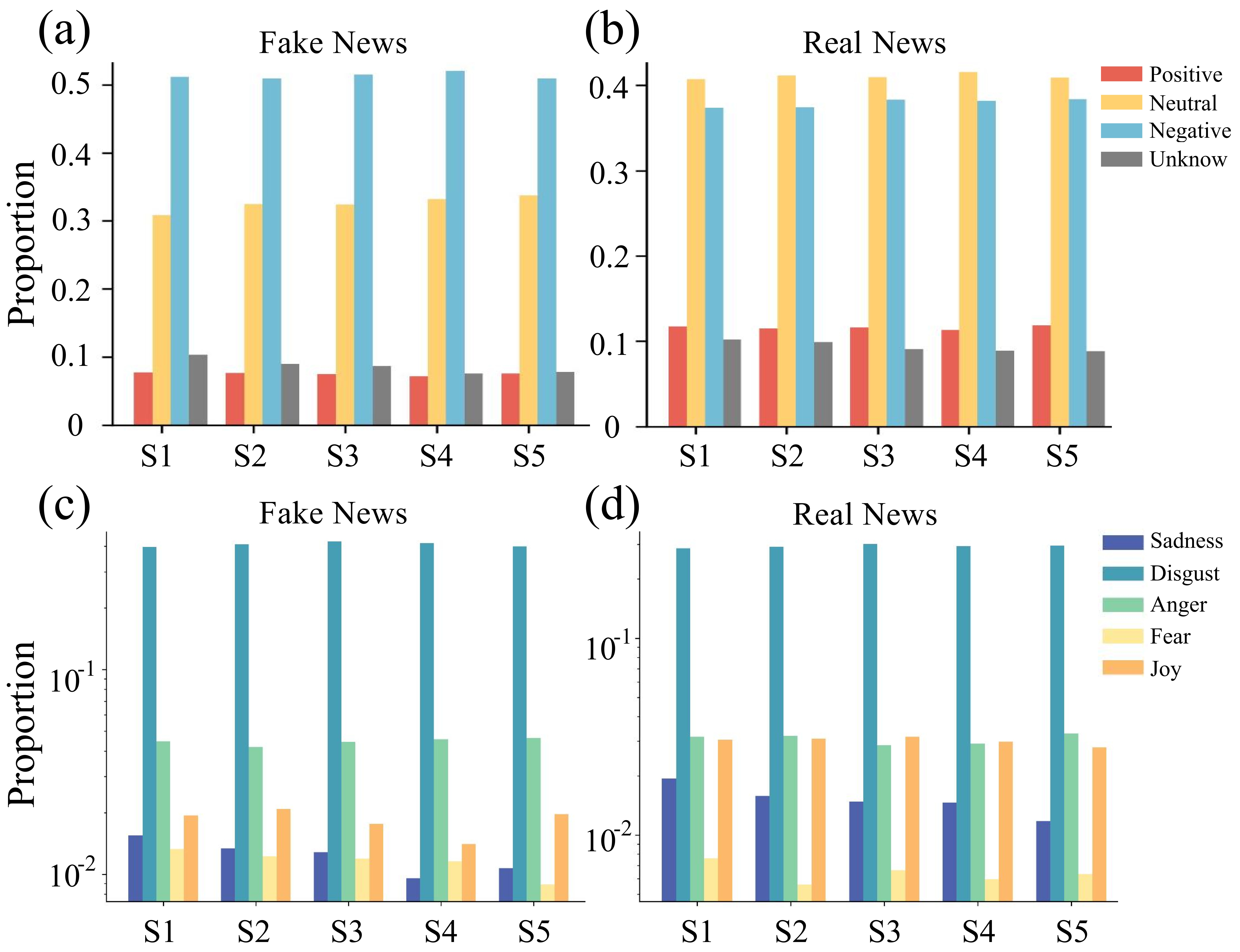} \caption{(a,b) Proportions of nodes with different sentiment polarities at each stage for fake and real news. (c,d) Proportions of nodes with different finer-grained emotions at each stage for fake and real news. In the analysis, each comment tree is divided into five stages based on its size. The first stage (S1) corresponds to the point when the tree grows to $20\%$ of its maximum size, the second stage (S2) spans from $20\%$ to $40\%$ growth, and so on.} \label{Fig:4}
\end{figure}

\subsection{Analysis of the user networks}

Finally, we investigate how agents interact from the perspective of user networks, which can be extracted from the aforementioned comment trees. A key concern here is that some users may appear in multiple comment chains, engaging in various discussions and maintaining interactions with different users. To address this, we merge different comments made by the same user into a single node (comments with unknown sentiment are removed). In the user network, each node represents a distinct user, and each edge denotes a comment-based interaction between users. The sentiment polarity of a user is calculated by averaging the polarities of the comments made by that user. Thus, unlike the comment trees, the polarity of a node in the user network cannot be represented by discrete integers. To facilitate analysis, we categorize nodes into three equal groups: $G_1$---nodes with negative sentiment, with polarity in the range $[-1,-1/3)$; $G_2$---nodes with neutral sentiment, with polarity in the range $[-1/3, 1/3]$; and $G_3$---nodes with positive sentiment, with polarity in the range $(1/3, 1]$. 

Figure 5a,c illustrates the average proportion of neighbors with sentiment $y$ pointing to a node with sentiment $x$ in the user network for fake and real news, where $x,y\in\{G_1,G_2,G_3\}$. Figure 5b,d shows the same results in the corresponding null models, where the sentiment polarities of nodes are randomly shuffled as described in Sec. A. We observe that, users with similar sentiments (particularly for negative and positive sentiments) tend to aggregate more tightly compared to the scenario where sentiment polarities are randomly mixed. These results align with previous findings on comment trees.

\begin{figure}
\includegraphics[width=1.0\linewidth]{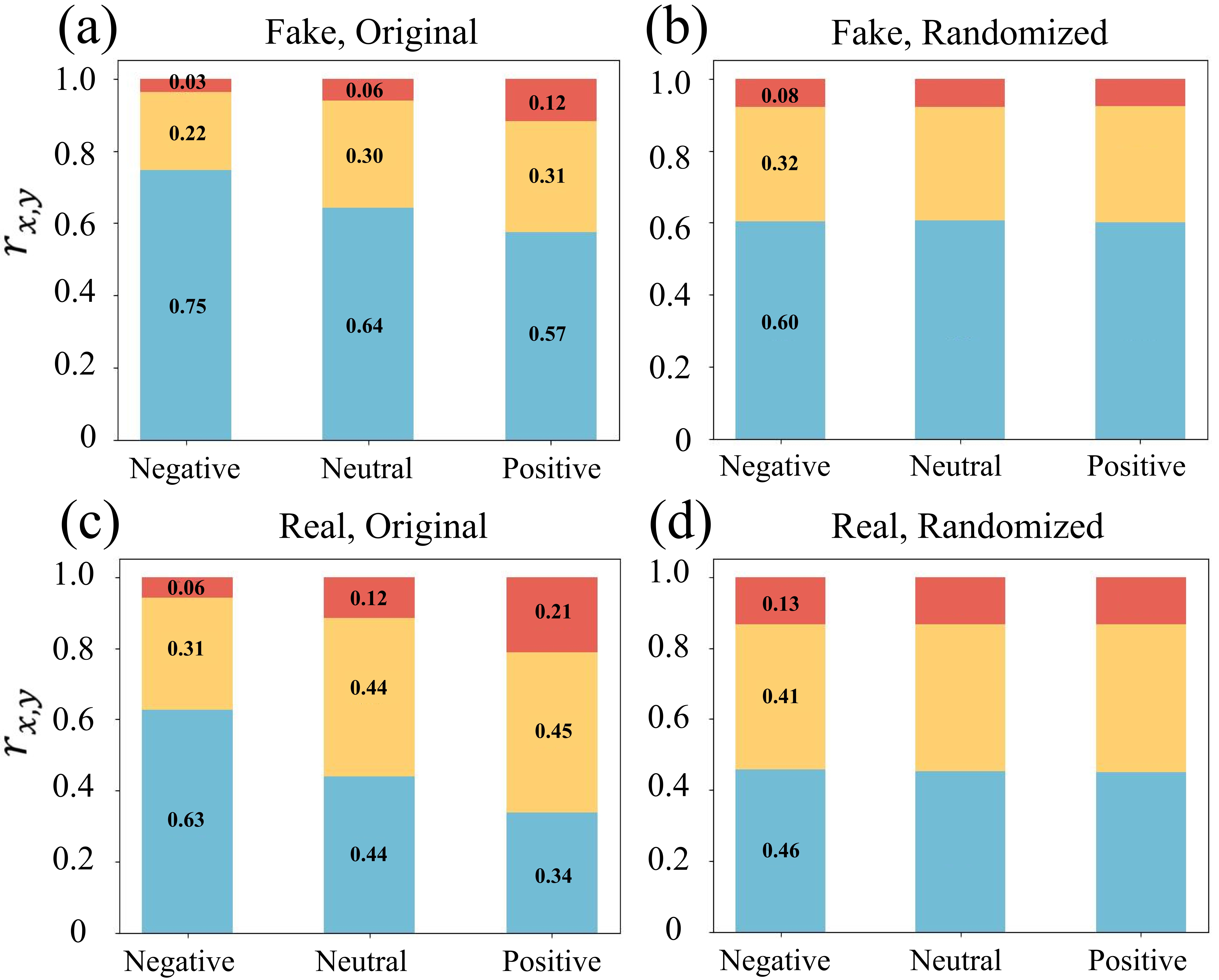} \caption{(a, c) $r_{x,y}$ for both fake and real news in user networks, where $x$ and $y$ could each denote negative, neutral, or positive sentiment. (b, d) $r_{x,y}$ in the corresponding null models, where the sentiment polarities of nodes in the user network are randomly shuffled (results averaged over $100$ realizations). The colors blue, yellow, and red represent negative, neutral, and positive sentiments, respectively.} \label{Fig:5}
\end{figure}

\section{Discussion}
In this paper, we have investigated the interaction patterns of agents in the comment sections of fake and real news. By analyzing data from Reddit, a large online social platform, we have identified a potential trend indicating that nodes in the fake news comment trees are more likely to form groups (defined as star networks consisting of several nodes). Moreover, we found that, compared to fake news, where users generate more negative sentiment, those discussing real news are more likely to exhibit neutral and positive sentiments. Nodes with similar sentiments also seem to cluster more tightly than expected, compared to a scenario where sentiments are randomly mixed. This may provide an evidence that emotions are contagious through pairwise interactions in social networks. In addition, from a dynamic perspective, we have demonstrated that, unexpectedly, the distribution of sentiment polarity in comment trees for both fake and real news stabilizes early on, remaining almost unchanged in the later stages.

It should be noted that this study has some limitations. First, our focus is on the Reddit platform. It remains unclear whether users on other social media platforms exhibit similar behavior, necessitating further research across different social platforms. Second, the data volume is limited; more data on fake news is required to reinforce the findings presented in this paper. Third, in reality, multiple social variables may influence the spread of fake news, such as agent competence \cite{Franceschi2022a,Franceschi2022b}. Therefore, a more comprehensive investigation is needed that considers factors beyond just node sentiments.

Our study may provide valuable insights into distinguishing between fake and real news on social media. Furthermore, it should be emphasized that while sentiment analysis is considered a key feature in discerning fake information in many machine learning models \cite{Setty2020,Ruchansky2017,Guo2019,Hamed2023,Alonso2021}, it is largely ignored in current theoretical modeling of social contagion processes \cite{Ruan2018,Iacopini2019,Ruan2015,Arruda2020,Watts2002}. Our work may help bridge this gap and enhance the understanding of certain social contagion dynamics in the real world.

\noindent Ethics. This work did not require ethical approval from a human subject or animal welfare committee.

\noindent Data accessibility. All source code required to replicate the results are publicly available at Zenodo \cite{code2024}.

\noindent Funding statement. This work was supported in part by the Key R\&D Program of Zhejiang under Grants 2022C01018 and 2024C01025, by the National Natural Science Foundation of China under Grants 62103374 and U21B2001.

\noindent Authors' contributions. Z.R. conceived the research project. K.Z., J.N. and S.P. performed research. Z.R., K.Z. and S.P. analyzed the results. Z.R. wrote the paper. All authors reviewed and approved the manuscript.

\noindent Conflict of interest declaration. We declare we have no competing interests.

\noindent Acknowledgements. We are thankful to the reviewers for their valuable and constructive comments.

\end{document}